\documentclass{mem}
\usepackage{natbib}\usepackage{txfonts}\usepackage{balance}
\usepackage{graphicx}
\usepackage[a4paper]{hyperref}
\idline{75}{282}
\begin{document}
\def\teff{$T\rm_{eff }$}
\def\kms{$\mathrm {km s}^{-1}$}

\title{
Red giant variables: OGLE--II and MACHO
}

   \subtitle{}

\author{
L.L.. \,Kiss\inst{1}
\and P. Lah\inst{2}
          }

\offprints{L.L. Kiss}

\institute{
School of Physics A28, University of Sydney, NSW 2006, Australia,
\email{laszlo@physics.usyd.edu.au}
\and
Research School of Astronomy \& Astrophysics,
Australian National University, Canberra, Australia
}

\authorrunning{L.L. Kiss \& P. Lah}

\titlerunning{Red giant variables: OGLE-II and MACHO}

\abstract{
We review the recent impact of microlensing projects on our
understanding of pulsating red giant stars.  Discussed are red giant stars'
pulsation properties (period--luminosity relations, period changes,
mode switchings), Red Giant Branch pulsations, metallicity effects and the 
use of red giant variables to explore galactic structure.
}
\maketitle{}

\section{Introduction}

The major problem with observational studies of Mira and semiregular stars has been the 
 long time scale of variability.  Since the
typical pulsation periods range from tens to hundreds of days, very few
long-term monitoring surveys existed in the pre-microlensing era. The lack of
extensive data and the lack of luminosity information for galactic
stars prevented real breakthroughs in understanding the general properties of
red giant oscillations. The current paradigm was born with the seminal works of
Wood et al.\ (1999) and Wood (2000), in which MACHO (Alcock et al. 2000) 
observations of red giant variables in the Large Magellanic Cloud (LMC) were
analysed in the period-luminosity (P--L) plane. The main results were summarized
in Wood (2000): {\it (i)}~there are five distinct period-luminosity relations
for red giant variables, of which four contain AGB stars; {\it (ii)}~Miras and
some low-amplitude semiregulars are fundamental mode pulsators (sequence C); 
{\it (iii)}~other semiregular stars can pulsate in the second and third
overtone modes (sequences A and B); {\it (iv)}~there are (first) red giant branch
eclipsing binaries with their own P--L relation (sequence E); {\it (v)}~a large
fraction of stars shows long-secondary periods, whose origin is still ambiguous 
(sequence D; for updates see Wood, these proceedings).  

Another major step in the field followed the publication of the OGLE-II (Udalski
et al. 1997) database of variable stars in the Magellanic Clouds (Zebrun et al.
2001).  This new interest in pulsating red giant stars has led to significant 
progress in several areas:

\begin{itemize}

\item pulsation properties

\item the presence of distinct evolutionary states

\item possible metallicity effects

\item red giant variables as probes of galactic structure

\end{itemize}

In the following sections we summarize how this progress has affected
our current view of Mira and semiregular variables.

\section{Pulsation properties}

\subsection{New features in the P--L plane}

\begin{figure*}[t!]
\begin{center}
\resizebox{6cm}{!}{\includegraphics[clip=true]{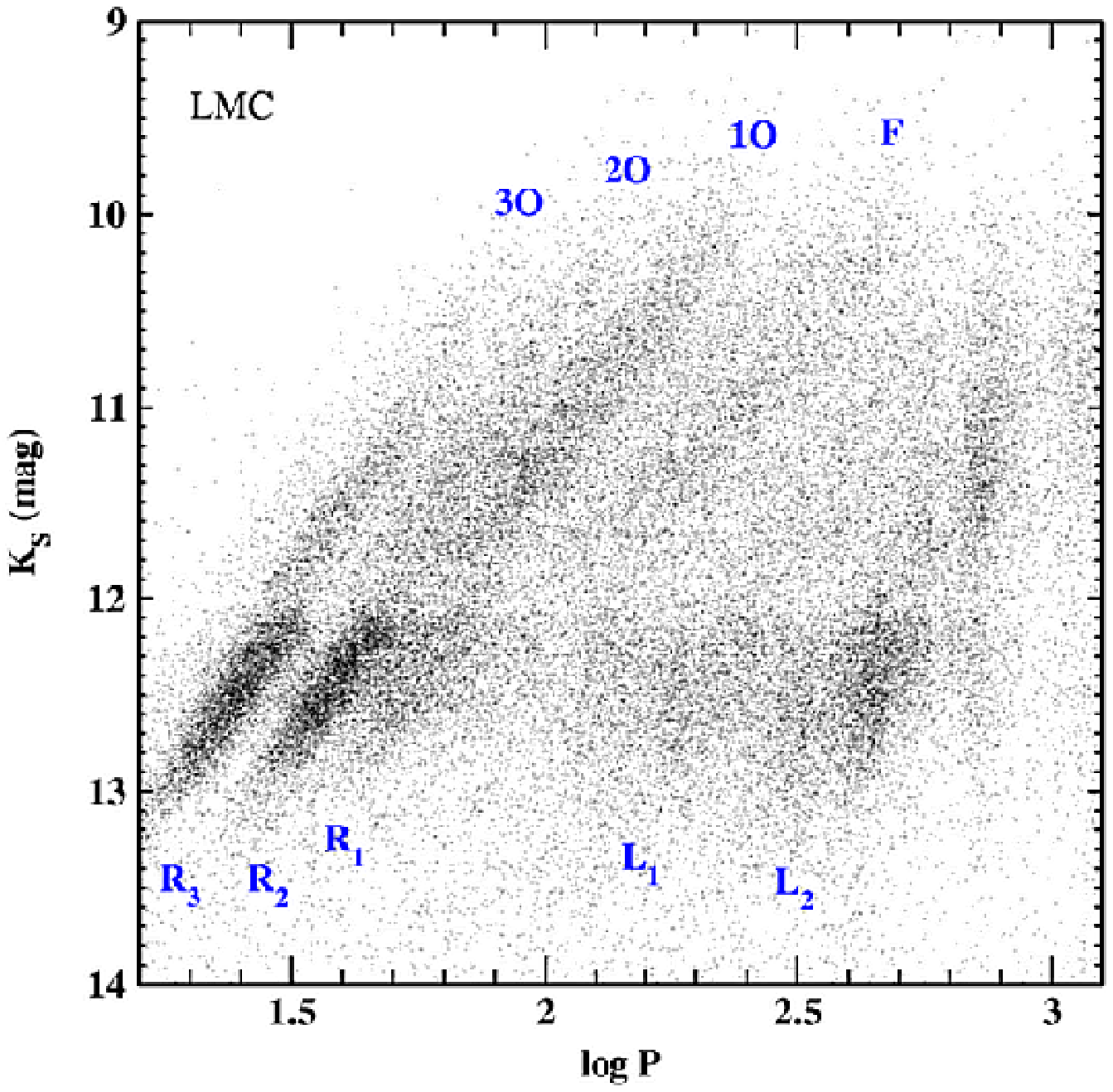}}
\resizebox{6cm}{!}{\includegraphics[clip=true]{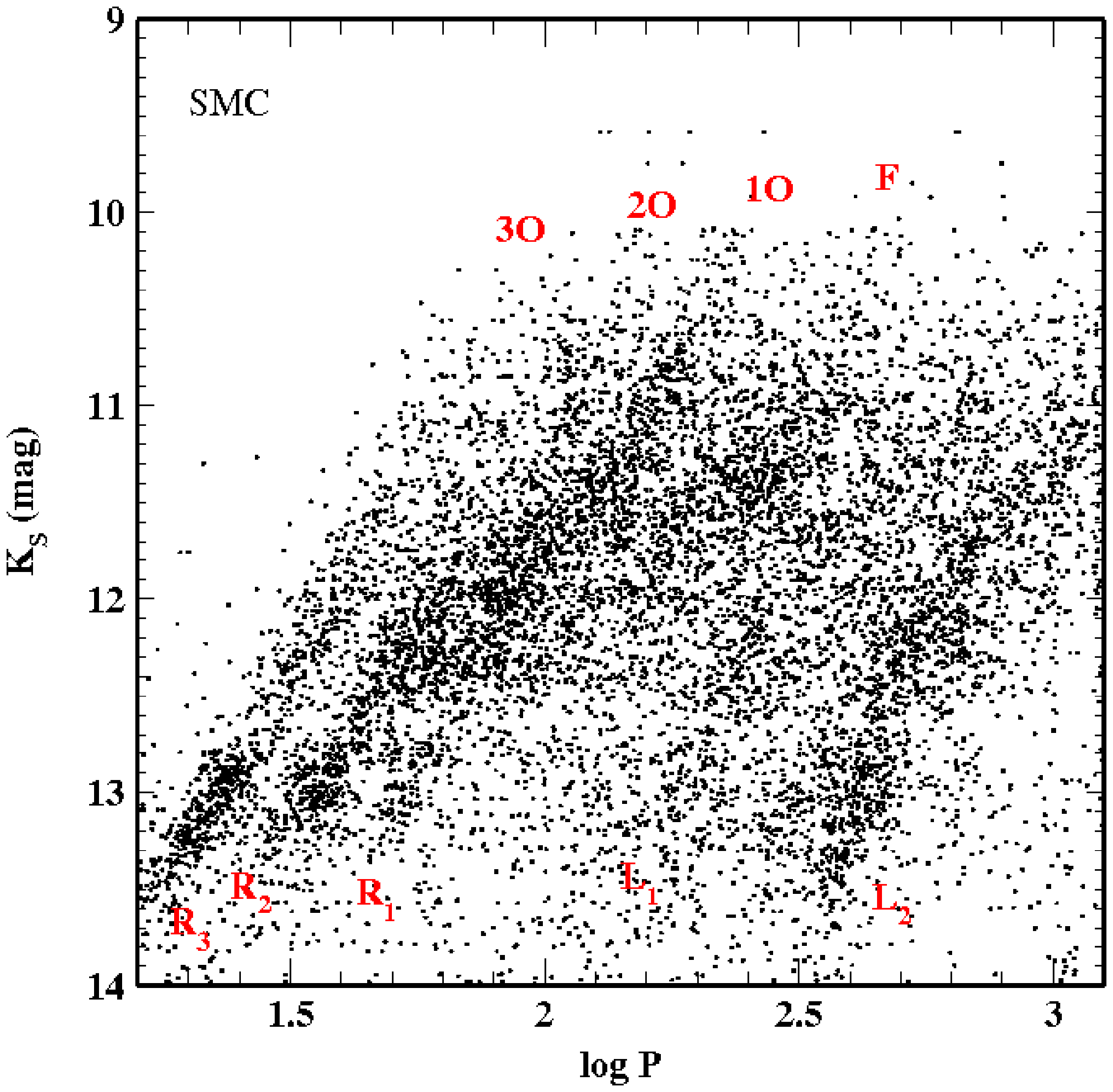}}
\caption{\footnotesize
Period-luminosity relations of variable red giants in the Magellanic Clouds,
using the OGLE-II and 2MASS data (Kiss \& Bedding 2003, 2004). Note the
difference in the luminosity extents of ridges R$_2$ and R$_3$. 
}
\label{pls}
\end{center}
\end{figure*}

The new large samples of variable red giants have improved our understanding of their properties.  
Using OGLE data from both Magellanic Clouds,  Kiss \& Bedding (2003, 2004), Ita et
al.\ (2004a\&b) and Soszynski et al.\ (2004) revealed many new
features in the period--luminosity plane. Groenewegen (2004) restricted his study to a 
sample of spectroscopically confirmed M-, S- and C- stars in the LMC and SMC OGLE data, 
while Fraser et al.\ (2005) analysed MACHO stars in the LMC only; SMC
C-stars were studied by Raimondo et al. (2005). An extensive Bulge study was
done by Wray et al. (2004).  The number of stars studied ranged from about 1,000 
(Raimondo et al.\ 2005) up to 20,000-24,000 (Kiss \& Bedding 2003, Fraser et
al. 2005).  The original Wood (2000) half-square degree sample in the
LMC contained only $\sim$800 stars.  The more than an order of magnitude increase 
in the number of variable red giants available for study has revealed new P--L sequences 
and sharp boundaries for the sequences. What originally seemed to be five sequences 
turned out to be an overlap of at least eight different P--L ridges, corresponding 
to both different modes of pulsations and distinct populations of stars. 

Fig.\ \ref{pls} shows the P--L relations from multiperiodic light curve fits of 
variable red giants in the LMC and SMC (Kiss \& Bedding 2003, 2004). The same
parallel P--L sequences can be seen in both galaxies, which means that overall
pulsation properties do not depend strongly on metallicity. The sharp
discontinuity at $K_S\approx12.05$ (LMC) and $K_S\approx12.70$ (SMC) is exactly
at the tip of the Red Giant Branch (TRGB).  This suggests that low-amplitude and
short-period stars below this edge belong to the shell hydrogen burning (first) Red
Giant Branch.  Further arguments for this interpretation are presented in the next Section.
The classical view of Mira and semiregular stars have them located on
the Asymptotic Giant Branch (AGB).  These stars below the TBRG are a whole new family of 
previously unknown oscillating stars.   They are short-period ($P<50$ days) and
low-amplitude ($A_I<0.02$ mag) variables, which largely bridge the gap between the classical semiregular
stars of early M spectral types and the very low-amplitude pulsating K-giants
(Edmonds \& Gilliland 1996).  Variable stars of this new type have also been found 
in the Galactic Bulge (Wray et al. 2004).  Their full potential for asteroseismology has yet
to be explored, including the identification of the excitation mechanisms in these
stars (``Mira-like'' vs. ``solar-like''  oscillations, Dziembowski et al.
2001).

The current sharpest view of the P--L sequences was presented by Ita et al.
(2004a\&b), who determined single periods for $\sim$9,000 stars in the LMC and
$\sim$3,000 stars in the SMC. Apart from reducing the effects of spurious
periods, they used SIRIUS $JHK$ magnitudes that go much deeper than the 2MASS
magnitudes.  Multiperiodicity is a common feature of these variable stars
and in most cases, secondary periods fall very close to the other
P--L sequences. It is multimode pulsations that are largely
responsible for the rich structure seen in Fig.\ \ref{pls}.

\begin{figure}[t!]
\resizebox{\hsize}{!}{\includegraphics[clip=true]{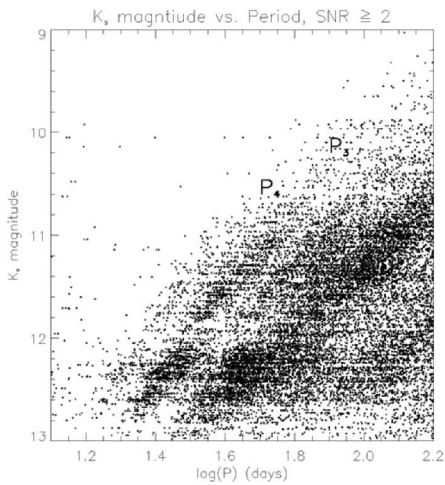}}
\caption{\footnotesize
The new shortest period P--L sequence of red giants in the LMC (marked 
P$_4$)
}
\label{p4}
\end{figure}

Recently, Soszynski et al.\ (2004) combined OGLE-II data with OGLE-III
observations to create a dataset that covers a time span comparable to the MACHO database.
They found multiperiodicity for the vast majority of the 15,400 small amplitude 
variable red giants in the LMC (and 3,000 stars in the SMC) 
and also that about 30\% of the sample exhibited two modes closely spaced in their
power spectrum. This is likely to indicate non-radial oscillations that is otherwise
hidden by the horizontal scatter of the P--L sequences. Soszynski et al. (2004) also 
noticed a previously overlooked high-order overtone sequence with
periods of 8 to 50 days, sitting on the left-hand side of the ridges R$_3$ and
3O in Fig.\ \ref{pls} (A$^+$ and A$^-$ in Ita et al. 2004a). This new sequence
is also seen in the publicly available MACHO data  (Fig.\ \ref{p4}), if all
frequencies with signal-to-noise ratios (SNR) greater than 2.0 are included in
the successive prewhitening steps. This new P--L sequence pushes up the
empirical limit for the acoustic cut-off frequency that traces the properties
of the outermost atmospheric regions.   This provides an important clue for the 
production of realistic stellar models.

\subsection{Mode switching and period change}

Mode switching in pulsating red giant stars was a rarely known phenomenon in
the  pre-microlensing era, because its detection requires very long
observational records. Convincing evidence was found in a few cases from many
decades of visual observations (e.g. Cadmus et al. 1991).  These results suggested 
the complex dynamics of these oscillations but the low number of
detected cases implies a low incidence rate of mode switching.

Groenewegen (2004) compiled a list of OGLE-II objects that were observed on
photographic plates between 1977 and 1984 by various groups. With a total of 370
stars, he compared historic periods with the more recent OGLE-II values and
found that about 10\% of the sample exhibited more than 10\% change in period
over 17 years. In most cases, different periods clearly  corresponded to
different P--L relations (see fig.\ 8 in Groenewegen 2004). Furthermore, he also
identified three stars that were classified oxygen-rich in the 1970's and
carbon-rich in the 1990's, suggesting that they may have undergone a thermal
pulse in the last 20 years and dredged-up enough carbon to switch spectral
type.  

\begin{figure*}[t!]
\begin{center}
\resizebox{5.7cm}{!}{\includegraphics[clip=true]{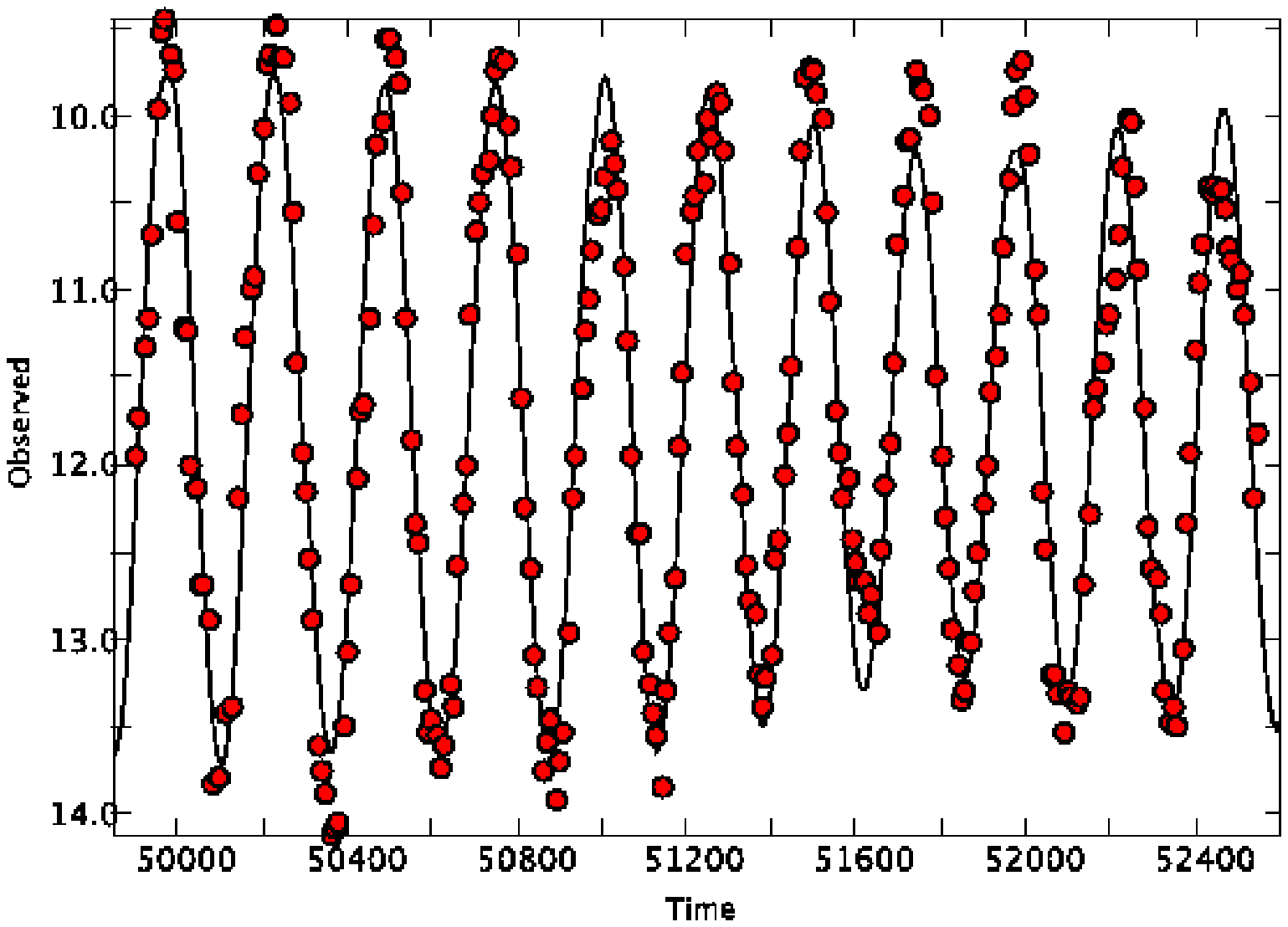}}
\resizebox{5.7cm}{!}{\includegraphics[clip=true]{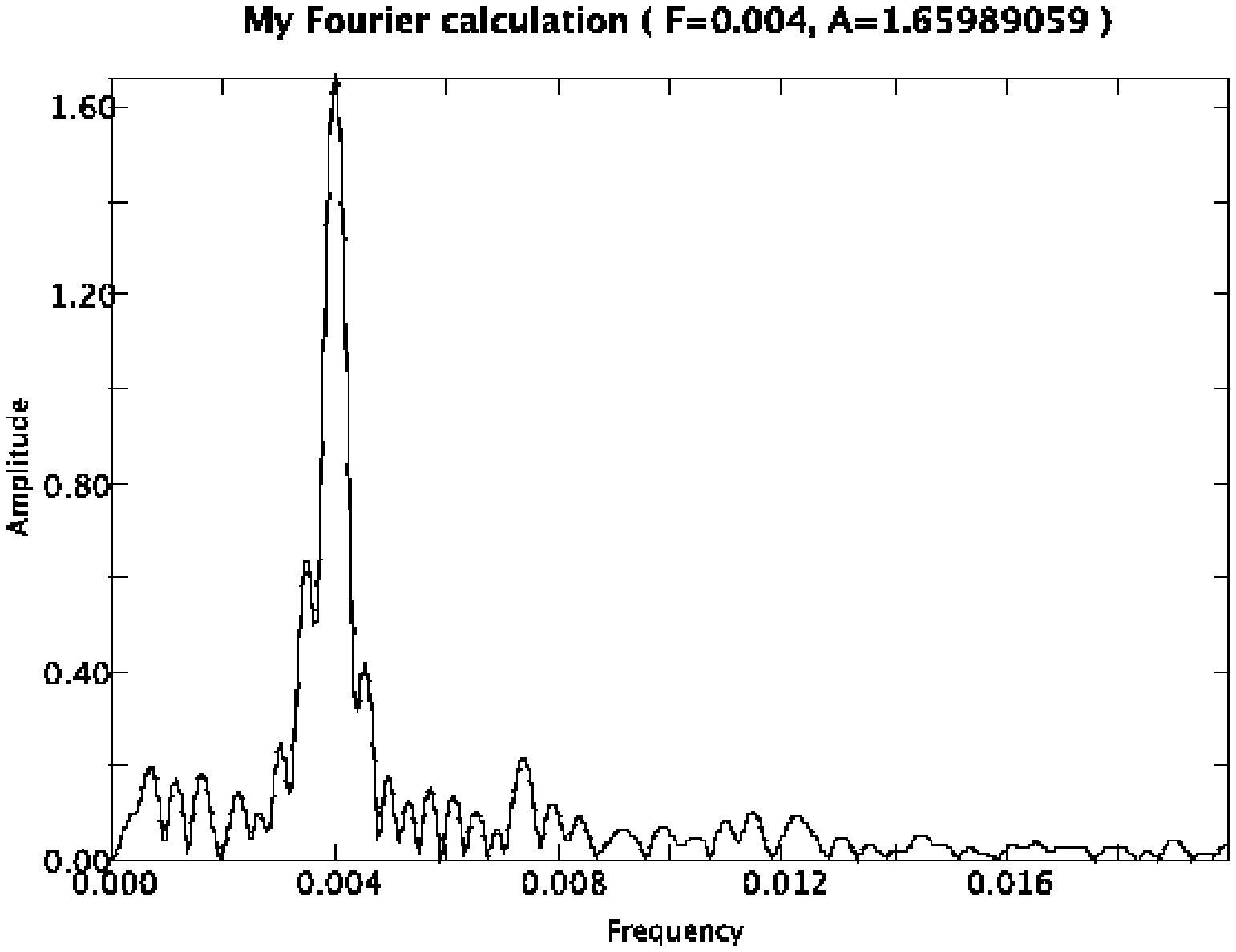}}
\resizebox{5.7cm}{!}{\includegraphics[clip=true]{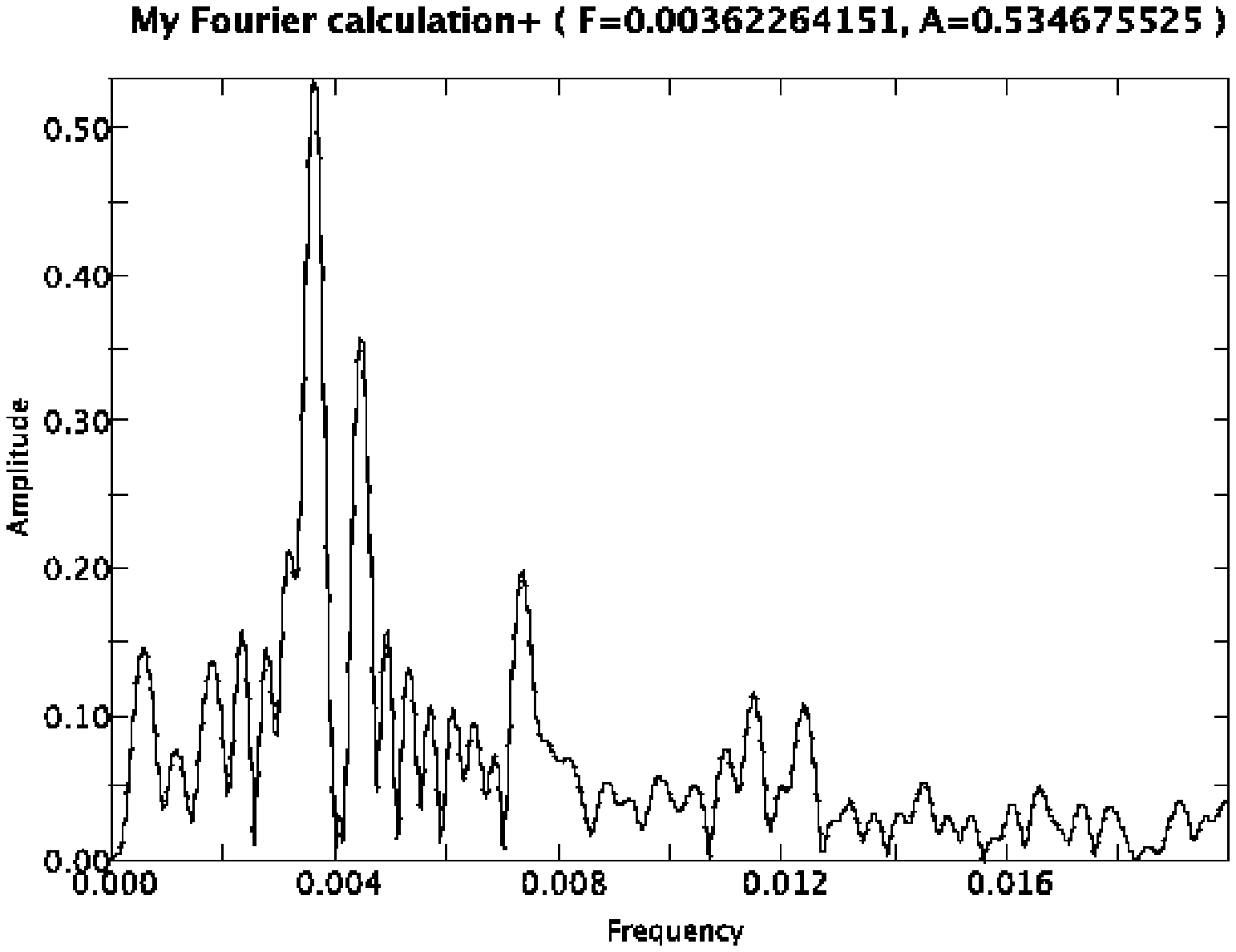}}
\resizebox{5.7cm}{!}{\includegraphics[clip=true]{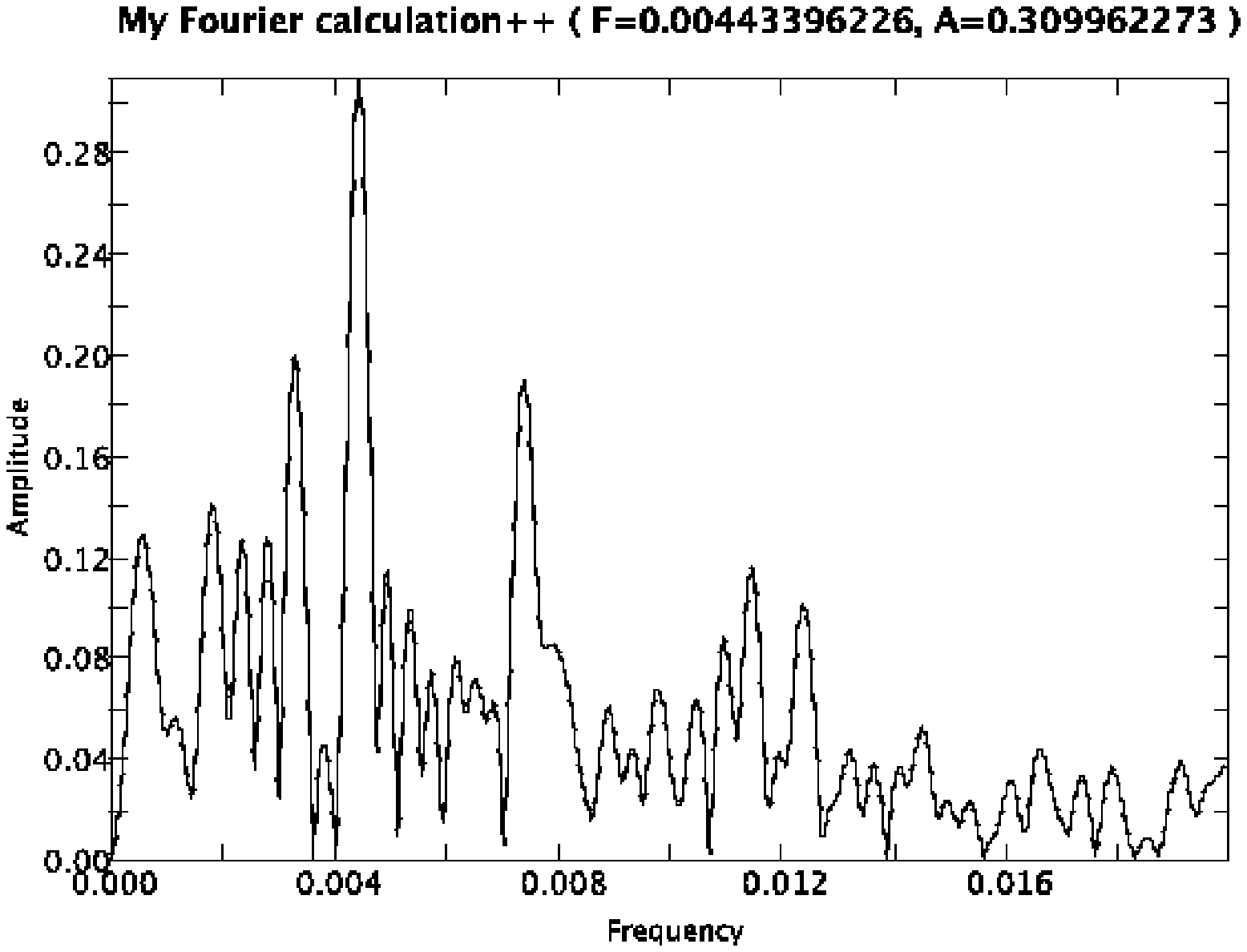}}
\caption{\footnotesize
Simulated T UMi-like light curve with the MACHO time span ({\it upper left panel})
and frequency spectra in three successive prewhitening steps ({\it upper right}:
initial spectrum; {\it lower right}: residual spectrum after the first prewhitening; 
{\it lower left}: and after the second prewhitening)
}
\label{tumimacho}
\end{center}
\end{figure*}

Theoretically, helium-shell flashes that occur on the AGB can be detected in
the form of very strong period changes in Mira variables (cf. Wood \& Zarro 1981).
The only problem is that Mira pulsations are intrinsically unstable in time and in many objects a
2--3\% period jitter or quasi-regular period shifts is seen.
Nevertheless, we know of a few galactic candidates for on-going thermal pulses.  The best 
example is T~Ursae~Minoris (Templeton et al. 2005). We examined whether the microlensing data might be able 
to identify further candidates for thermal pulses.  To do this we took two 
subsets of the visual light curve of T~UMi, one with
the OGLE-II time span and one with the MACHO time span.  Both covered the strongest period
change in that star (see more details on T~UMi in Szatm\'ary et al. 2003). We then
performed successive prewhitenings in the power spectrum and checked what
frequencies resulted from this procedure. 

The four years of OGLE-II would not have been enough to detect any
measurable period change in a star like T~UMi, because the period analysis
resulted in only the mean period (250 days) and its harmonics. However, the eight
years of the MACHO project would have allowed one to pick up the changing period
of the star: the highest peaks after successive subtractions remained close to 
the initial one, being a good indicator of strong period change 
(see Fig.\ \ref{tumimacho}).  We suggest that the existing MACHO data
(or the combined OGLE-II+OGLE-III data) could be used for selecting
candidates  for on-going thermal pulses.  Spectroscopic monitoring of these stars could
help us better understand this intriguing phase of late stellar evolution.

\section{RGB vs. AGB variables}

The existence of pulsations at the tip of the first red giant branch was
initially proposed by Ita et al. (2002). Kiss \& Bedding (2003) and subsequently
Ita et al. (2004a) suggested that short-period OGLE-II red giants below the
TRGB are at least partly RGB stars. Besides the sharp edge in  their luminosity
function, they found a measurable period shift within the two most populated
P--L sequences, which also occurred at the TRGB. This shift can be fully
explained by the mean temperature difference of RGB and AGB stars at the same
luminosity, which was the second argument for RGB pulsations. The third argument
was presented by Kiss \& Bedding (2004), who compared multicolour luminosity
functions of OGLE-II red giant variables in the LMC and SMC. The tip of the RGB,
as determined from the $IJHK$ luminosity functions, showed a colour and
metallicity dependence that is in excellent agreement with the empirical results
for globular clusters. Finally, Soszynski et al.\ (2004) have shown
quantitatively that the short-period  P--L sequences below the TRGB are in fact
a mixture of RGB and AGB variables, except the newly identified sequence (cf.
Fig.\ \ref{p4}) that contains second ascent AGB red giants.

These low-amplitude pulsations in RGB stars have a number of interesting
applications. They may explain the long-standing problem of
``velocity jitter'' in field and globular cluster red giants (Gunn \& Griffin
1979, Carney et al. 2003). This phenomenon manifests as a higher than expected
dispersion in radial velocity measurements, with the greatest velocity
variations seen within 0.5 mag (in V) of the TRGB. Although Carney et al. (2003)
argued against radial pulsations, we propose that these newly identified RGB
oscillations, that have their largest photometric amplitudes around the TRGB, can
cause this ``velocity jitter''. The multiperiodic nature of these pulsations
coupled with seemingly irregular behaviour (mode switching, sudden amplitude
changes) is the likely reason why existing radial velocity observations showed
only a random scatter.  The low-amplitude RGB pulsations may also be connected 
to the Ca II K emission line asymmetries seen in red giants (Smith \& Shetrone 2004).

\section{Metallicity effects}

\begin{figure}[t!]
\resizebox{\hsize}{!}{\includegraphics[clip=true]{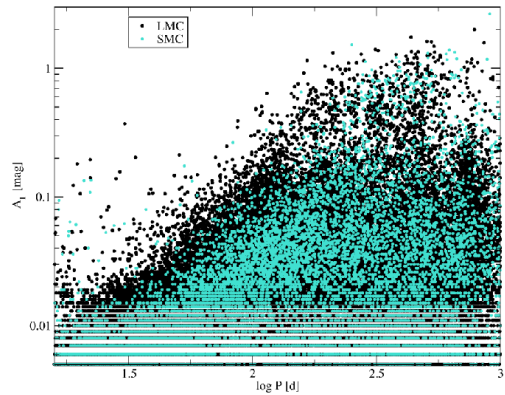}}
\caption{\footnotesize
OGLE-II I-band period--amplitude relations for the LMC (black dots) and the SMC
(light gray dots).}
\label{logp-amp}
\end{figure}

Using the large microlensing datasets it is possible to do a comparison of the
Magellanic Clouds and the Galactic Bulge looking for metallicity effects on red
giant pulsations and P--L relations. A zero-point differences of the order of
0.1 mag were found by Glass \& Schultheis (2003) and Ita et al.\ (2004a), while
slight changes in the slopes for each relation were also  detected by Ita et
al.\ (2004a) and Lah et al.\ (2005). Furthermore, Schultheis et al.\ (2004) 
made a comprehensive comparison of the Magellanic Clouds and the Bulge and found
that the proportion of stars that vary decreases at lower metallicities and the
minimum period associated with a given amplitude increases.  At any given
period, the most metal-rich stars in the Bulge  have the highest photometric
amplitude and the most metal-poor stars in the SMC  have the lowest. This agrees
with expectations as the optical amplitudes will in general be smaller at lower
metallicity as they have weaker bands of highly sensitive molecules like TiO.
This would favour small amplitude variability in metal-poor environments and
explains the smaller fraction of large amplitude variables in the SMC compared
to the LMC and the Galactic Bulge.  Plotting over 60,000 periods and amplitudes 
for $\sim$23,000 stars in the LMC and 10,000 period-amplitude pairs for
$\sim$3,200 stars in the SMC reveals this general trend (Fig.\ \ref{logp-amp}):
at any given period, the maximum I-band amplitude of variation is roughly twice
as large in  the LMC as in the SMC.  

Other differences between the red giant pulsations in the three galaxies include 
different luminosity ranges for each P--L sequences (Wray et al. 2004, Schultheis 
et al. 2004; also visible in Fig.\ \ref{pls}) and the different fraction of multiperiodic stars with
certain period ratios (Soszynski et al. 2004). The emerging view of metallicity
effects is very complex which offers strong empirical constraints for theory.

\section{Galactic structure from pulsating red giants}

\begin{figure}[t!]
\resizebox{\hsize}{!}{\includegraphics[clip=true]{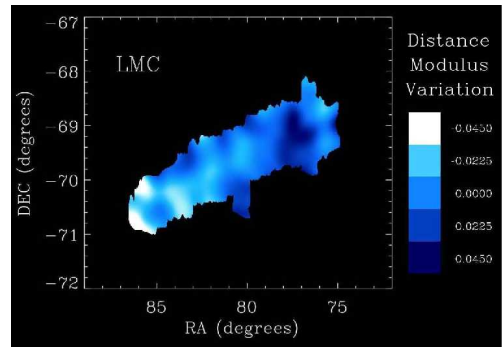}}
\caption{\footnotesize
A 3-D representation of the LMC. The lighter regions are closer to us and the
darker regions are further away.}
\label{lmc3d}
\end{figure}

The width of the P--L relations is affected by many different factors, one of which
is the line-of-sight distance variations of the galaxy in question. Lah et al.\
(2005) presented the first attempt to use red giant P--L relations to constrain
the three-dimensional structure of the Magellanic Clouds. For this work it is was 
assumed that for a star, at a given period, the vertical difference between its 
observed K magnitude and the mean linear P--L relation can be taken as the
star's distance modulus relative to the average distance to the host
galaxy. A distance modulus from one star will have great uncertainty, but from 
thousands of stars a clear trend produced by the inclination of the galactic disk 
can be seen.  In Fig.\ \ref{lmc3d} we show the structure 
traced by 4,276 RGB pulsators in the LMC bar, which is in excellent agreement 
with the view based on other distance
indicators (Cepheids, red clump stars, etc.): the east side of the LMC is closer
than the west, while the inclination angle of the LMC bar is $\sim$30$^\circ$ and
the distance variation across the face of the LMC is measured at $\sim$2.4 kpc.

Recent Bulge studies include Wray et al. (2004) and Groenewegen \& Blommaert 
(2005). Both papers discussed the geometry of the Galactic Bar from pulsating
red giants, making use of the excellent statistical properties of the OGLE-II
sample. Although red giant P--L relations are not as tight as that of the
Cepheids, it is evident that the large number of Mira and
semiregular stars makes them valuable distance indicators.    

\begin{acknowledgements}
L.L. Kiss has been supported by the Australian Research Council and a University
of Sydney Postdoctoral Research Fellowship. Fig.\ \ref{tumimacho} has been
produced with {\sc Period04}.
\end{acknowledgements}

\bibliographystyle{aa}

\end{document}